\documentstyle[preprint,aps,prc,rotate,array]{revtex}
\begin{document}
\input{psfig}
\input{epsf}
\def\Im{\mbox{\sl Im\ }}
\def\pd{\partial}
\def\oln{\overline}
\def\olft{\overleftarrow}
\def\ds{\displaystyle}
\def\bgreek#1{\mbox{\boldmath $#1$ \unboldmath}}
\def\sla#1{\slash \hspace{-2.5mm} #1}
\newcommand{\bra}{\langle}
\newcommand{\ket}{\rangle}
\newcommand{\vep}{\varepsilon}
\newcommand{\met}{{\mbox{\scriptsize met}}}
\newcommand{\lab}{{\mbox{\scriptsize lab}}}
\newcommand{\cm}{{\mbox{\scriptsize cm}}}
\newcommand{\mcal}{\mathcal}
\newcommand{\Del}{$\Delta$}
\newcommand{\g}{{\rm g}}
\long\def\Omit#1{}
\long\def\omit#1{\small #1}
\def\beq{\begin{equation}}
\def\eeq{\end{equation} }
\def\bea{\begin{eqnarray}}
\def\eea{\end{eqnarray}}
\def\eqref#1{Eq.~(\ref{eq:#1})}
\def\eqlab#1{\label{eq:#1}}
\def\figref#1{Fig.~\ref{fig:#1}}
\def\figlab#1{\label{fig:#1}}
\def\tabref#1{Table \ref{tab:#1}}
\def\tablab#1{\label{tab:#1}}
\def\secref#1{Section~\ref{sec:#1}}
\def\seclab#1{\label{sec:#1}}
\def\VYP#1#2#3{{\bf #1}, #3 (#2)}  
\def\NP#1#2#3{Nucl.~Phys.~\VYP{#1}{#2}{#3}}
\def\NPA#1#2#3{Nucl.~Phys.~A~\VYP{#1}{#2}{#3}}
\def\NPB#1#2#3{Nucl.~Phys.~B~\VYP{#1}{#2}{#3}}
\def\PL#1#2#3{Phys.~Lett.~\VYP{#1}{#2}{#3}}
\def\PLB#1#2#3{Phys.~Lett.~B~\VYP{#1}{#2}{#3}}
\def\PR#1#2#3{Phys.~Rev.~\VYP{#1}{#2}{#3}}
\def\PRC#1#2#3{Phys.~Rev.~C~\VYP{#1}{#2}{#3}}
\def\PRD#1#2#3{Phys.~Rev.~D~\VYP{#1}{#2}{#3}}
\def\PRL#1#2#3{Phys.~Rev.~Lett.~\VYP{#1}{#2}{#3}}
\def\FBS#1#2#3{Few-Body~Sys.~\VYP{#1}{#2}{#3}}
\def\AP#1#2#3{Ann.~of Phys.~\VYP{#1}{#2}{#3}}
\def\ZP#1#2#3{Z.~Phys.~\VYP{#1}{#2}{#3}}
\def\ZPA#1#2#3{Z.~Phys.~A~\VYP{#1}{#2}{#3}}
\def\half{\mbox{\small{$\frac{1}{2}$}}}
\def\quarter{\mbox{\small{$\frac{1}{4}$}}}
\def\nn{\nonumber}
\newlength{\PicSize}
\newlength{\FormulaWidth}
\newlength{\DiagramWidth}
\newcommand{\vslash}[1]{#1 \hspace{-0.5 em} /}
\def\olaf{\marginpar{Mod-Olaf}}
\def\her{\marginpar{$\Longleftarrow$}}
\def\bel{\marginpar{$\Downarrow$}}
\def\abo{\marginpar{$\Uparrow$}}

\tighten


\title{Low-energy Compton scattering on the nucleon and sum rules}

\author{S. Kondratyuk}
\address{
TRIUMF, 4004 Wesbrook Mall, Vancouver, British Columbia, Canada V6T 2A3}
\author{O.\ Scholten}
\address{
Kernfysisch Versneller Instituut, University of Groningen, 9747 AA
Groningen, The~Netherlands}

\date{\today}

\maketitle

\begin{abstract}

The Gerasimov-Drell-Hearn and Baldin-Lapidus sum rules are evaluated in 
the dressed K-matrix model for
photon-induced reactions on the nucleon.  
For the first time the sum $\alpha+\beta$ of the electric and magnetic
polarisabilities and the forward spin polarisability $\gamma_0$ are
explicitly calculated in two alternative ways -- from the sum rules 
and from the low-energy
expansion of the real Compton scattering amplitude -- within 
the {\em same} framework.
The two methods yield
compatible values for $\alpha+\beta$ but differ somewhat for $\gamma_0$. 
Consistency between the two ways of determining the polarisabilities is a
measure of the extent to which basic symmetries of the model are obeyed.


\end{abstract}



\section{Introduction} 
\seclab{intro}

The interest in the low-energy photon-nucleon interaction has been
renewed in recent years. A number of experiments 
have been carried out \cite{Mac95} to study real Compton 
scattering at low energies -- where the amplitude
is parametrised by nucleon polarisabilities -- and up to the second
resonance region. 
The new global data fit (last Ref.\ in~\cite{Mac95}) has yielded 
the electric and magnetic polarisabilities 
$\alpha=12.2 \pm 0.7[\mbox{stat+syst}]$ and 
$\beta=1.7 \mp 0.7[\mbox{stat+syst}]$, 
and the forward spin-polarisability 
$\gamma_0=-1.87 \pm 0.18[\mbox{stat+syst}]$ has been extracted
from the recent 
measurements~\cite{Ahr01} of the total photoabsorption cross section.
(We use the standard units $10^{-4}$ fm$^3$ 
for $\alpha$ and $\beta$ and $10^{-4}$ fm$^4$ for $\gamma_0$.)
Dispersion theory has been successfully
utilised to extract nucleon polarisabilities from data with a minimum of model
assumptions.
We quote results of two recent dispersion calculations:
$\alpha=11.9$, $\beta=1.9$ (first Ref.\ in~\cite{Bab98}),
$\gamma_0=-0.8$ (second Ref.\ in~\cite{Bab98}).
General relations between
polarisabilities and total photoabsorption cross sections are provided by the
Baldin-Lapidus (BL), Gerasimov-Drell-Hearn (GDH)  
and other sum rules \cite{Bal60}, which have been evaluated in several
meson-exchange models \cite{Dre01}. 

Nucleon polarisabilities have been calculated in
various dynamical frameworks 
from a low-energy expansion of the Compton scattering amplitude. 
Next-to-leading order calculations in the chiral perturbation theory
yielded for the scalar polarisabilities
$\alpha=10.5 \pm 2.0$, $\beta=3.5 \mp 3.6$~\cite{Ber94}, and 
$\gamma_0=-3.9$~\cite{Kum00} (or $-3.8$~\cite{Ji00}).
The polarisabilities calculated in the dressed K-matrix model  
\cite{Kon01} are $\alpha=12.1$, $\beta=2.4$ and $\gamma_0=2.4$.

The low-energy expansion and the sum rules
represent two alternative ways of calculating nucleon polarisabilities.
This paper is the first attempt to compare these two methods in 
the {\em same} dynamical framework. 
We use the dressed K-matrix model (expounded in \cite{Kon01}), which
is a relativistic, unitary, crossing symmetric and gauge-invariant approach 
with certain analyticity constraints  
incorporated through a dressing procedure for propagators and vertices.
This dressing with meson loops up to infinite order 
is the central element of our approach. 
The model describes pion-nucleon scattering, pion photoproduction and Compton
scattering at both intermediate and low energies.

We consider two particular combinations of the polarisabilities
which can be extracted from the low-energy expansion of the forward scattering
Compton amplitude,
on the one hand, and calculated from sum rules, on the other. These are 
the sum $\alpha + \beta$ of the
electric and magnetic polarisabilities
and the forward spin polarisability $\gamma_0$.
The purpose of this work is to focus attention on the general question of
agreement between the polarisabilities extracted
from the low-energy expansion of the amplitude and those obtained from the sum
rules. 
Since the properties of unitarity, causality (analyticity of the amplitude), 
crossing symmetry and gauge
invariance put stringent constraints on both low-energy expansion and sum rules,
a discrepancy between the two ways of determining the polarisabilities is
a rather direct measure of the extent to which the basic symmetries are 
obeyed in the model. We find agreement in
the leading order terms related to the anomalous magnetic moment (the GDH sum
rule) and $\alpha + \beta$ (the BL sum rule), but some violation in the next
order related to $\gamma_0$.
We will show the effects of certain model assumptions. 

\section{Polarisabilities from the low-energy Compton scattering}
\seclab{pol_cs} 

The forward Compton
scattering amplitude is expanded in a small photon energy $\omega$ as
\beq 
T(\omega)=\overrightarrow{\epsilon}' 
\! \cdot \! \overrightarrow{\epsilon}\!
\left[ - {e^2 \over {4 \pi m}} + (\alpha+\beta) \omega^2 \right]
\!+\! i \overrightarrow{\sigma} 
(\overrightarrow{\epsilon}' \! \times \! 
 \overrightarrow{\epsilon})\!
\left[ -{{e^2 \kappa^2} \over {8 \pi m^2}} \omega + \gamma_0 \omega^3 
\right] \!+\! {\mathcal{O}}(\omega^4), 
\eqlab{low}
\eeq
where $\overrightarrow{\epsilon}$ and
$\overrightarrow{\epsilon}'$ are the polarisations of the initial and
final photons, respectively, and $\overrightarrow{\sigma}$,
$m$, $e$ and $\kappa$
are the spin vector, mass, charge and anomalous
magnetic moment of the nucleon. The constant and linear terms are
given by the Born contribution, and the model-dependent 
subleading terms are determined by the polarisabilities 
$\alpha$, $\beta$ and $\gamma_0$. 

The polarisabilities calculated in our model 
from the low-energy expansion of the
Compton amplitude are given in the columns labelled ``LE" in 
\tabref{polar}, where the original calculations are presented in the three upper
rows (the lower rows will be explained below). 
How the polarisabilities are affected by increasing the ``amount of dressing"
is best explained in terms of the kernel of the calculation, the K-matrix.  
The bare calculation corresponds to a K-matrix being the sum
of the tree-level s- and u-channel diagrams in which bare vertices and free
propagators are used. This approximation is used in traditional K-matrix models
\cite{Kor98}, where only the pole parts of the loop integrals 
(and no principal-value parts) are included in the
T-matrix by iterating the K-matrix, hence
the maximal violation of analyticity in this calculation. 
The second row in \tabref{polar} is a calculation 
in which analyticity is partially restored since
the K-matrix is now constructed out of dressed propagators and vertices
(2- and 3-point functions), thereby including the principal-value parts of a 
large class of loop diagrams. 
This restoration of analyticity is not complete, 
however, since the 4-point $\gamma \gamma N N$
contact term is not dressed as the 2- and 3-point functions,
only its longitudinal part being uniquely determined 
from gauge invariance.
This deficiency is 
mitigated in the full calculation, where the transverse part of the 
$\gamma \gamma N N$ vertex includes a ``cusp" term~\cite{Kon01} 
taking into account the principal-value part of the 
loop diagram where the two photons couple to the intermediate pion.           
     
\section{Polarisabilities from the sum rules}
\seclab{pol_sr}

The sum rules~\cite{Bal60} relate the low-energy observables to
the photoabsorption cross sections 
corresponding to the total angular momenta $1/2$ and $3/2$.  
The GDH sum rule involves the proton anomalous magnetic moment $\kappa=1.79$, 
\beq
{m^2 \over 2 \pi e^2} \int_{\omega_{th}}^{\infty}\!d\omega \, 
\frac{\sigma_{1/2}-\sigma_{3/2}}{\omega} = -\frac{\kappa^2}{4},
\eqlab{gdh}
\eeq
where $\omega_{th}$ is the pion-production threshold energy.
We also evaluate the BL sum rule for 
$\alpha+\beta$ and the sum rule for $\gamma_0$, which can be written,
respectively, as  
\beq
{1 \over 4 \pi^2} \int_{\omega_{th}}^{\infty} \! d \omega \,
\frac{\sigma_{1/2}+\sigma_{3/2}}{\omega^2} = \alpha+\beta, \;\;\;
{1 \over 4 \pi^2} \int_{\omega_{th}}^{\infty} \! d \omega \,
\frac{\sigma_{1/2}-\sigma_{3/2}}{\omega^3} = \gamma_0.
\eqlab{blsr_gamsr}
\eeq 

The integrand of the GDH sum rule \eqref{gdh} is shown in \figref{gdh_int},
where we display the results of the fully dressed (solid line) 
and bare (sparse-dotted line) calculations 
as well as the contribution of $\pi N$ intermediate states 
to the full calculation (dense-dotted line).
The data points are from~\cite{Ahr01}, and 
the result of the unitary isobar model (first in Refs.~\cite{Dre01}) 
is shown for comparison (dashed line).
While in good overall agreement with experiment, 
our results deviate from it for photon energies above 700 MeV.
The GDH and BL integrals 
are shown as functions of the upper limit of
integration in Figs.~\ref{fig:gdh_sum} and \ref{fig:ab_sum}, respectively.
Convergence is not achieved below 
1 GeV of photon energy (we remark that
the data neither support nor rule out the absence of convergence at
these energies). 
The inelastic contributions (beyond the $\pi N$ intermediate
states) become significant above 450 MeV and bring the calculation
to a better agreement with the data.
The spin-polarisability sum rule 
converges by 1 GeV, as \figref{gdh_g0sum} shows. 
Moreover, this convergence is essentially
due to the $\pi N$ channel, a feature
consistent with the strong suppression of the integrand
of the $\gamma_0$ sum rule at higher energies.            

\section{Comparing the low-energy and sum-rule values of polarisabilities}
\seclab{compar}

The dynamical model 
used in this calculation and the sum rules are based on the 
same general principles: unitarity, crossing and CPT symmetry, gauge invariance
and analyticity, with the model approximations described above regarding
analyticity.   
Thus, if the calculated T-matrix obeyed the property of analyticity exactly,
the polarisabilities extracted from the low-energy expansion would be
equal to those obtained from the sum rules.

The polarisabilities obtained from the sum rules 
are given in the columns labelled ``SR" in \tabref{polar}.
The values of the polarisabilities are taken at the energies where the
corresponding sum rules converge, $\approx 1.5$ GeV for $\alpha+\beta$,
and $\approx 1$ GeV for $\gamma_0$.   
The dressing has a significant influence on the
low-energy polarisabilities and a minor effect on the sum rules.  
This trend is especially pronounced for the spin polarisability $\gamma_0$. 
The additional dressing of the $\gamma \gamma N N$
contact term has a large effect on the 
low-energy values of polarisabilities but is negligible for the sum rules.
This is because the cusp contact term affects only the $f_{EE}^{1-}$ 
partial wave in the region of the pion-production threshold \cite{Kon01}, 
while the sum rules integrate the contributions of all partial waves. 
The effects of restoration of analyticity on the amplitude are most pronounced
at lower energies.
At order $\sim \omega^2$, i.e.\ for $\alpha + \beta$, the
agreement between the low-energy and sum-rule results is improved by the
dressing of the 2- and 3-point functions and is further refined by the 
included contribution of the dressed 4-point function.
A problem occurs at third order, i.e.\ for $\gamma_0$,
where there is a disagreement 
between the low-energy and sum-rule values.                           
   
In the formulation of the model attention has been mainly focused on the 
consistent dressing of the nucleon. 
Therefore, the disagreement between the sum rules and the low-energy expansion
is likely to be related to treating other degrees of freedom not on the same
footing with the nucleon. We will
discuss two possible extensions of the dressing procedure, 
concentrating on the spin polarisability since here the discrepancy is
most conspicuous.

\noindent
(1) The resonances beyond the $\Delta$ have not been
included in the dressing equations, partly for simplicity and partly 
because one expects the associated violation of analyticity to be small.
To investigate this more
explicitly, we did a calculation in which
the only resonance kept in the K-matrix was the $\Delta$. 
In this case the low-energy values of polarisabilities are not notably affected,
and the sum rule for the spin polarisability is given by the dashed line in
\figref{gdh_g0sum}. 
It is seen that including the higher resonances in the dressing
would not eliminate the disagreement between the low-energy expansion 
and the sum rule.

\noindent
(2) In the present version of the model, the $\Delta$ self-energy is computed
in a one-loop approximation only, and dressing of the $\pi N \Delta$ vertex is
not considered. However, as was shown in Ref.~\cite{Kon01}, the multi-loop
corrections tend to strongly enhance the scalar part of the nucleon self-energy.
Therefore, we did an exploratory calculation wherein the $\Delta$
self-energy 
$\Sigma_{\Delta}(p)=A_{\Delta}(p^2)\vslash{p}+B_{\Delta}(p^2)m_{\Delta}$
was modified in a similar way:
its scalar part $B_{\Delta}(p^2)$ was increased below the nucleon mass, 
i.e.\ at $p^2<m^2$, from its original value of $\approx 1.28$ to 
$\approx 1.35$. This amount of increase was sufficient to match the
low-energy and the sum-rule values of the spin polarisability.
The polarisabilities obtained in this exploratory calculation
are given in the lower three rows of \tabref{polar}.
Since the $\Delta$ self-energy is altered only
far from the resonance mass, the amplitudes of pion and photon scattering
on the proton, and hence the sum rules, are unaffected.

In summary, 
we compared proton polarisabilities calculated
from the low-energy Compton scattering with those obtained from
the corresponding sum rules.
Although the BL sum rule appears
not converged by an energy of 1 GeV (which we consider an upper limit for the
validity of the model), at this energy the sum rule agrees with the 
low-energy polarisability.
Similarly, the GDH sum rule at $\approx 1$ GeV is consistent with
the proton anomalous magnetic moment. 
These agreements show
in particular that the model obeys the essential causality 
constraints.
A discrepancy appears only at higher -- third -- order in photon energy,
which is related to the spin polarisability.



We thank Peter Grabmayr, Harold Fearing and Andrew Lahiff for
discussions. 
S.K. is supported by a grant from the Natural Sciences and
Engineering Research Council of Canada. The work of O.S. is 
part of the research program of the ``Stichting voor
Fundamenteel Onderzoek der Materie'' (FOM) with financial support
from the ``Nederlandse Organisatie voor Wetenschappelijk
Onderzoek'' (NWO).



\begin{table}
\caption[t1]{Proton polarisabilities 
calculated from the low-energy Compton amplitude \eqref{low} and
from the sum rules Eqs.~(\ref{eq:blsr_gamsr}).
The different ingredients of the dressing are explained in 
\secref{pol_cs}. The three lower rows are the results of the exploratory 
calculation described in \secref{compar}.
}
\begin{center}
\begin{tabular}{c c c c c} \hline
 & \multicolumn{2}{c}{$\;\;\;\;\;\;\;\;\;\; \alpha+\beta$} & 
 \multicolumn{2}{c}{$\gamma_0$} \\ \hline
 & LE  & \hspace*{-5mm} SR  & \hspace*{2mm} LE  & \hspace*{1mm} SR \\ \hline
Bare  & $15.8\!+\!1.4\!=\!17.2$ & \hspace*{-5mm} 14.5 & 
\hspace*{1mm} $-0.9$ & $-1.2$ \\ 
2,3-pt.funct. & $\;\, 8.9\!+\!2.4\!=\!11.3$ & \hspace*{-5mm} 13.8 & 
\hspace*{1mm} $-0.1$ & $-0.7$  \\ 
Full dressing & $12.1\!+\!2.4\!=\!14.5$ & \hspace*{-5mm} 13.8 & 
\hspace*{1mm} $\;\;\, 2.4$ 
& $-0.7$ \\      
\hline
 &  \multicolumn{1}{c}{\it Modified $\Delta$ self-energy} 
 \\ \hline
Bare  & $15.8\!+\!1.4\!=\!17.2$ & \hspace*{-5mm} 14.5 & 
\hspace*{1mm} $-0.9$ & $-1.2$ \\ 
2,3-pt.funct. & $\;\; 8.9\!+\!1.5\!=\!10.4$ & \hspace*{-5mm} 13.8 & 
\hspace*{1mm} $-3.3$ & $-0.7$  \\ 
Full dressing & $  12.1\!+\!1.6\!=\!13.7$ & \hspace*{-5mm} 13.8 & 
\hspace*{1mm} $ \,-0.7$ & $-0.7$ \\      
\hline
\end{tabular}
\end{center}
\tablab{polar}
\end{table}


\begin{figure}
\centerline{{\epsfxsize 7.2cm  \epsffile[35 55 590 415]{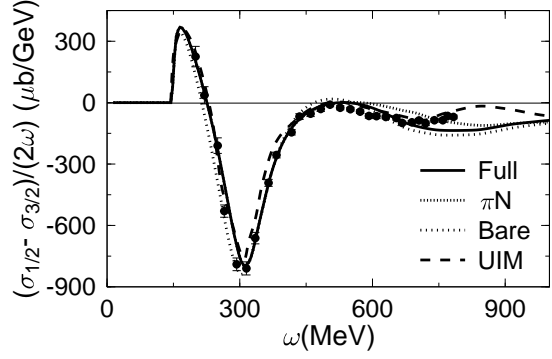}}}
\caption[f2]{The energy dependence  
of the integrand of the GDH sum rule \eqref{gdh}. 
The curves are explained in \secref{pol_sr}.
The data in all figures are from~\cite{Ahr01}.
\figlab{gdh_int}} 
\end{figure}

\begin{figure}
\centerline{{\epsfxsize 7.2cm  \epsffile[0 55 555 415]{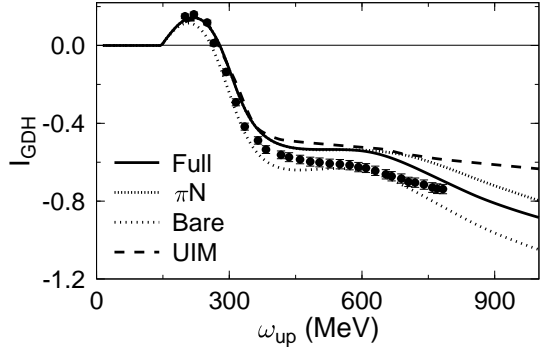}}}
\caption[f2]{The GDH sum rule \eqref{gdh}  
as a function of the upper limit of
integration. 
\figlab{gdh_sum}} 
\end{figure}

\begin{figure}
\centerline{{\epsfxsize 7.2cm  \epsffile[0 55 555 415]{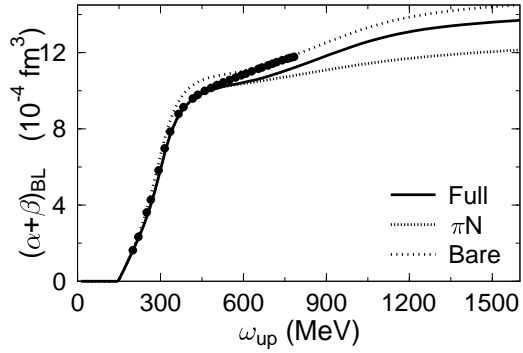}}}
\caption[f2]{The BL sum rule,
first of Eqs.~(\ref{eq:blsr_gamsr}),
as a function of the upper limit of integration.  
\figlab{ab_sum}} 
\end{figure}

\begin{figure}
\centerline{{\epsfxsize 7.2cm  \epsffile[35 55 590 415]{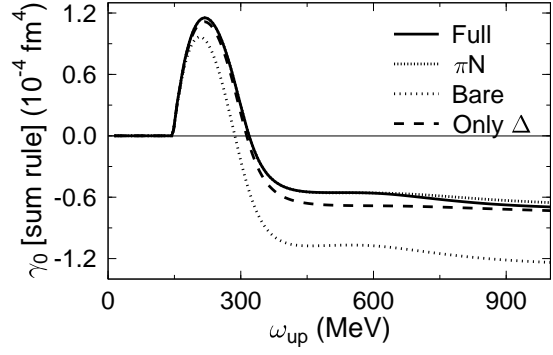}}}
\caption[f2]{The $\gamma_0$ sum rule, 
second of Eqs.~(\ref{eq:blsr_gamsr}), 
as a function of the upper limit of integration. 
The dashed line is obtained in the full calculation
in which the only retained resonance is the $\Delta$.    
\figlab{gdh_g0sum}} 
\end{figure}

\end{document}